\documentclass[conference]{IEEEtran}
\usepackage{graphicx}
\usepackage{subfig}
\usepackage{fixltx2e}
\usepackage{url}
\usepackage{mathtools}
\usepackage[T1]{fontenc}
\usepackage{multirow}
\usepackage{mathtools}

\begin{document}

\title{Diversity, Productivity, and Growth of Open Source Developer Communities}

\author{%
\makebox[.5\linewidth]{Qingye Jiang}\\School of Information Technologies\\The University of Sydney\\Sydney NSW 2006, Australia\\Email: qjiang@ieee.org
\and \makebox[.5\linewidth]{Young Choon Lee}\\Department of Computing\\Macquarie University\\Sydney NSW 2109, Australia\\Email: young.lee@mq.edu.au\\
\and \makebox[.5\linewidth]{Joseph G. Davis}\\School of Information Technologies\\The University of Sydney\\Sydney NSW 2006, Australia\\Email: joseph.davis@sydney.edu.au
\and \makebox[.5\linewidth]{Ablert Y. Zomaya}\\School of Information Technologies\\The University of Sydney\\Sydney NSW 2006, Australia\\Email: albert.zomaya@sydney.edu.au
}

\maketitle
\thispagestyle{plain}
\pagestyle{plain}

\begin{abstract}
The open source development model has become a paradigm shift from traditional in-house/closed-source software development model, with many successes. Traditionally, open source projects were characterized essentially by their individual volunteer developers. Such tradition has changed significantly with the participation of many organizations in particular. However, there exists a knowledge gap concerning how open source developer communities evolve. In this paper, we present some observations on open source developer communities. In particular, we analyze git repositories of 20 well-known open source projects, with over 3 million commit activities in total. The analysis has been carried out in three respects, productivity, diversity and growth using the Spearman's rank correlation coefficient, diversity index and the Gompertz/logistic curves, respectively. We find out that (a) the Spearman's rank correlation coefficient between active contributors and commit activities reveals how changes in the size of the developer community impacts the productivity of the community; (b) the diversity index of an open source developer community reveals the structure of the community; and (c) the growth of open source developer communities can be described using different phases of growth curves as in many organic matters.
\end{abstract}

\begin{IEEEkeywords}
open source, community, diversity, productivity, growth curve
\end{IEEEkeywords}

\section{Introduction}
\label{sec:intro}

``Community'' refers to a group of people having particular characteristics in common, which can be either a certain territorial and graphical location, or a set of interests and skills \cite{gusfield1975community}. In the case of open source projects, an open source project's developer community includes the developers contributing code to the project. During the past two decades, the information and communications technology industry has witnessed a paradigm shift from traditional in-house software development model to an increased reliance on the open source development model. 

In the early days of the open source movement, open source was often described as a volunteer community working in an unorganized way in return for primarily intrinsic rewards \cite{raymond1999cathedral}. This is then followed by an increasing number of organizations participating in the open source movement with the clear goal of commercial success \cite{nagy2010organizational} \cite{capiluppi2012exploring}. In particular, when a proprietary solution becomes successful in the market, there quickly appear one or more open source followers with similar features, such as MapReduce \cite{mapreduce} and Hadoop \cite{web:hadoop}, and VMWare \cite{web:vmware} and Xen \cite{xen}. When multiple open source projects coexist in the same market segment, these projects compete with each other for market share, e.g., cloud computing projects such as OpenStack \cite{sefraoui2012openstack} , CloudStack \cite{sabharwal2013apache} \cite{kumar2014apache}, Eucalyptus \cite{eucalyptus} and OpenNebula \cite{milojivcic2011opennebula}. The result of such competition is often reflected in the size and commit activity in their respective developer communities. 

It is commonly agreed that the growth of an open source developer community is associated with the degree of adoption of the open source project \cite{bonaccorsi2003open} \cite{crowston2003defining} \cite{english2007identifying}. However, there exists a knowledge gap concerning how open source developer communities evolve over time. Many founding organizations of open source projects are aware of the fact that making the source code publicly available alone is not sufficient to attract individuals and organizations to join the project. In many open source projects, there exists a role called Community Manager \cite{goth2005open} \cite{jensen2007role} \cite{michlmayr2009community} whose goal is to promote the project in a broader developer community, and to encourage individuals and organizations to join the project. As the number of contributing organizations increases, it is vaguely perceived that the diversity of the developer community also increases \cite{lerner2002some} \cite{crowston2003defining} \cite{meneely2009secure}. In practice, we all know that some projects are bigger or more productive than others. However, from a theoretical point of view, very little is known on how diversity affects the size and productivity of an open source developer community.

In this paper, we explore the following questions of theoretical and practical importance:

\begin{itemize}
	\item How does the size of an open source developer community affects the productivity of the community?
	\item How to measure the diversity of open source developer communities? How does diversity affects the productivity of the community?
    \item How to describe the growth of open source developer communities?
\end{itemize}

To answer these questions, we analyze the git repositories of 20 well-known open source projects in various disciplines (Table \ref{tbl:population}). These git repositories are up to date as of December 2015, with over 3 million commit activities. In particular, the analysis has been carried out in (a) the Spearman's rank correlation coefficient between the number of active contributors and the number of commit activities; (b) the diversity of the developer communities using diversity index; and (c) the growth of the developer communities using the Gompertz curve and the logistic curve.

We find out that Spearman's rank correlation coefficient between size and productivity reveals the relationship between the various contributing organizations in the same developer community. For all open source projects, there exist a core contributor group making the majority of the commit activities. Statistically, the contributions from the community members can be described by the power law distribution. When multiple contributing organizations coexist in the same open source project, the diversity index can be used to describe the structure of the developer community. The diversity index is effective in identifying projects dominated by a single organization and projects collectively developed by multiple organizations. The growth of an open source project is similar to the growth of bacteria in a nutrient-containing broth.

The rest of this paper is organized as follows. In Section \ref{sec:data}, we describe the acquisition of data used in this study. We present and discuss our observations on open source developer communities with respect to productivity, diversity and growth in Sections \ref{sec:productivity}, \ref{sec:diversity} and \ref{sec:growth}, respectively. Section \ref{sec:related} describes existing literature in open source community related researches. Our conclusions are presented in Section \ref{sec:conclusion}.

\section{Data Acquisition}
\label{sec:data}

We use the git repositories of the open source projects (Table \ref{tbl:population}) being studied as our data source. The ``git log'' command was used to extract information about commit activities, including date and time of the commits, as well as names and email addresses of the contributors. Several factors are considered when selecting the open source projects, including the number of contributors, the number of contributing organizations, and the volume of commit activities. The selected project must have over 100 unique contributors and 20 contributing organizations over its life span, plus an average of 100 commit activities in a month. Two exceptions being Eucalyptus and OpenNebula.

Unique developers are identified by their email addresses. When a developer commits code to a project in a particular month, he/she is considered an active contributor of the respective developer community in that month. However, the same developer might stop contributing code in other months, during which he/she should be considered as inactive.  A developer community is growing when the number of active contributors increases, and withering when the number of active contributors decreases. We are aware of the fact that a small number of developers do contribute code to the same open source project using multiple email addresses. Our manual checking indicates that this is not a common practice for the projects selected. Such practice does not significantly affect the results reported in this paper.

Similarly, active contributing organizations refers to organizations contributing code to a project in a particular month. We identify these organizations by the domain names in the contributor's email addresses. The organization that contributes the largest number of commits is said to be the primary contributing organization of the project. In general, there are three types of domain names observed from git logs. The first type is corporate domain names including companies such as intel.com and research institutes such as berkeley.edu. Contributors associated with corporate domain names usually work for the entity represented by the domain name. In this case, the commit activities are usually part of their job assignments, and represent the contribution from the entity they work for. When a contributor changes job he/she gets a new corporate email address, which correctly reflects the new entity he/she works for. The second type is virtual organization domain names such as apache.org and gnome.org. The contributors are affiliated with the virtual organization in some way, but usually work for other companies or research institutes. Certain open source projects prefer the usage of virtual organization email addresses. For example, most contributors for the Apache Hadoop project use apache.org email addresses. This makes it difficult to identify the actual organizations contributing to the project. For this reason, we exclude projects from this study when the primary contributing entity is a virtual organization. The third type is email service providers such as gmail.com and hotmail.com. In this case, the contributor chooses to contribute to the open source project as an individual.  

\begin{figure}[t!]
\centering
  \includegraphics*[width=8.3cm]{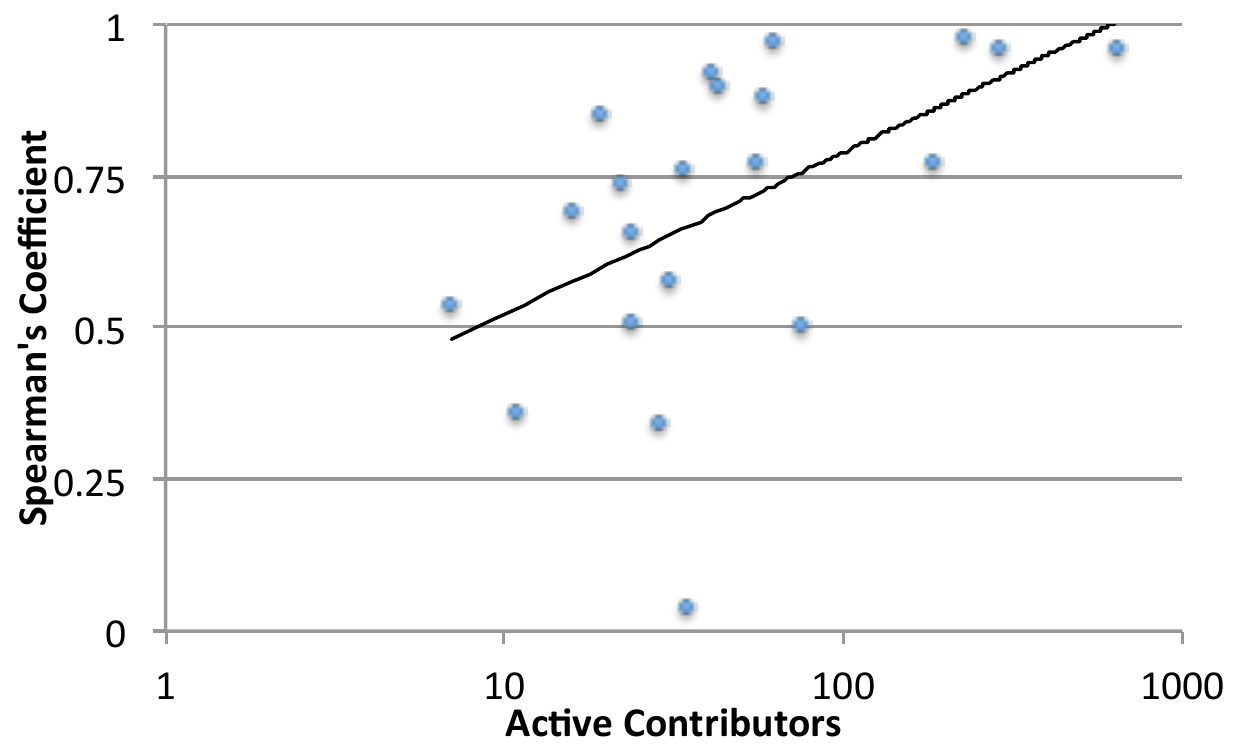}
  \caption{Community Size and Productivity.} 
  \label{fig:productivity} 
\end{figure}

\section{Productivity}
\label{sec:productivity}

In the git logs, each commit activity represents some changes in the code (additions or deletions). We use the number of commit activities to represent the \emph{productivity} of an open source project. For two open source projects with approximately the same number of active contributors, the one with more commit activities is said to be more productive than the other. For a particular open source project, as the number of active contributors increases, it is generally expected that the number of commit activities would increase accordingly. However, for different open source projects, the degree of such increase varies significantly. 

The relationship between the number of active contributors and the number of commit activities can be represented by the Spearman's rank correlation coefficient \cite{pirie1988spearman}, which is defined as the Pearson correlation coefficient between ranked variables. The Spearman's rank correlation coefficient is a nonparametric measure of statistical dependence between two variables. It represents how well the relationship between two variables can be described using a monotonic function. The Spearman's rank correlation coefficient ($\rho$) can be calculated by

\begin{equation}
\label{eq:pearson}
\rho = 1 - \frac{6\sum_{i=1}^{n}(X_{i} - Y_{i})^{2}}{n (n^{2} - 1)}
\end{equation}
\\where \textit{X} is an array representing the number of active contributors, \textit{Y} is an array representing the number of commit activities, and \textit{n} is the size of arrays \textit{X} and \textit{Y}.

\begin{table*}[t!]
\caption{Summary of Our Observation and Analysis.}
\label{tbl:population}
\centering
\begin{tabular}{|p{2.0cm}|p{1.8cm}|p{1.8cm}|p{1.8cm}|p{1.8cm}|p{1.8cm}|p{1.8cm}|} \hline
	Project & Total \newline contributors & Active \newline contributors & Monthly \newline commits & Active \newline organizations & Spearman's\newline coefficient & Diversity\newline index \\ \hline
	OpenNebula & 80 & 5 -- 10 & 50 -- 100 & 1 -- 5 & 0.64 & 1.11  \\ \hline
	Eucalyptus & 80 & 10 -- 40 & 100 -- 400 & 5 -- 10 & 0.54 & 1.83  \\ \hline
	glibc & 370 & 20 -- 40 & 100 -- 550 & 5 -- 25 & 0.54 & 1.83  \\ \hline
	CloudStack & 370 & 20 -- 70 & 200 -- 800 & 5 -- 20 & 0.04 & 2.14  \\ \hline
	Puppet & 530 & 20 -- 40 & 100 -- 300 & 8 -- 20 & 0.69 & 2.24  \\ \hline
	Gnome & 710 & 20 -- 40 & 200 -- 500 & 10 -- 20 & 0.96 & 2.52  \\ \hline
	MongoDB & 340 & 20 -- 45 & 300 -- 850 & 5 -- 10 & 0.88 & 1.76  \\ \hline
	Erlang &  370 & 30 -- 40 & 160 -- 360 & 10 -- 20 & 0.74 & 1.33  \\ \hline
	Python & 190 & 30 -- 45 & 300 -- 1000 & 15 -- 30 & 0.85 & 3.43  \\ \hline
	PHP & 590 & 30 -- 55 & 400 -- 1200 & 5 -- 25 & 0.34 & 1.08  \\ \hline
	Golang & 590 & 30 -- 60 & 250 -- 800 & 5 -- 20 & 0.90 & 1.51  \\ \hline
	Emacs & 730 & 30 -- 70 & 50 -- 1000 & 20 -- 50 & 0.66 & 2.02  \\ \hline
	binutils & 590 & 40 -- 60 & 300 -- 600 & 20 -- 35 & 0.58 & 3.54  \\ \hline
	MySQL &  360 & 50 -- 90 & 350 -- 700 & 2 -- 15 & 0.77 & 1.63  \\ \hline
	Apache Spark & 700 & 50 -- 130 & 150 -- 500 & 10 -- 50 & 0.92 & 4.18  \\ \hline
	Docker & 1040 & 60 --140 & 300 -- 450 & 30 -- 50 & 0.50 & 4.28  \\ \hline
	Webkit &  590 & 60 -- 200 & 1000--3200 & 5 -- 20  & 0.97 & 1.89  \\ \hline
	Mozilla & 4100 & 180 -- 460 & 2000--5300 & 80 -- 140  & 0.77 & 2.76  \\ \hline
	OpenStack & 3800 & 400 -- 700 & 2000--6000 & 100 -- 200 & 0.98 & 5.60  \\ \hline
	Linux Kernel & 15870 & 600 -- 1100 & 3800--6600 & 300 ~ 400 & 0.96 & 9.90  \\ \hline
\end{tabular}
\end{table*}

\begin{figure*}[t!]
\centering
 \subfloat[Gnome (\(\rho\)=0.96).]{ 
    \label{fig:subfig:spearman_gnome}  
    \includegraphics*[width=6cm]{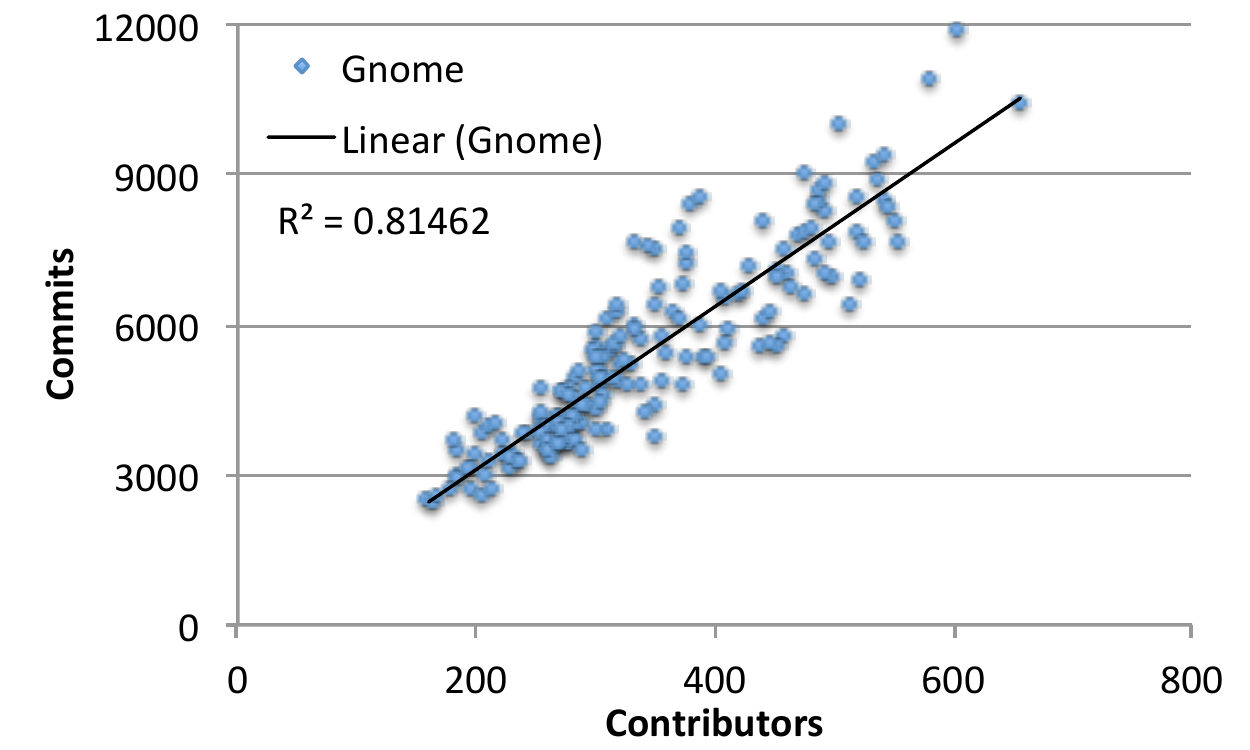}} 
  \subfloat[Mozilla (\(\rho\)=0.77).]{ 
    \label{fig:subfig:spearman_mozilla} 
    \includegraphics*[width=6cm]{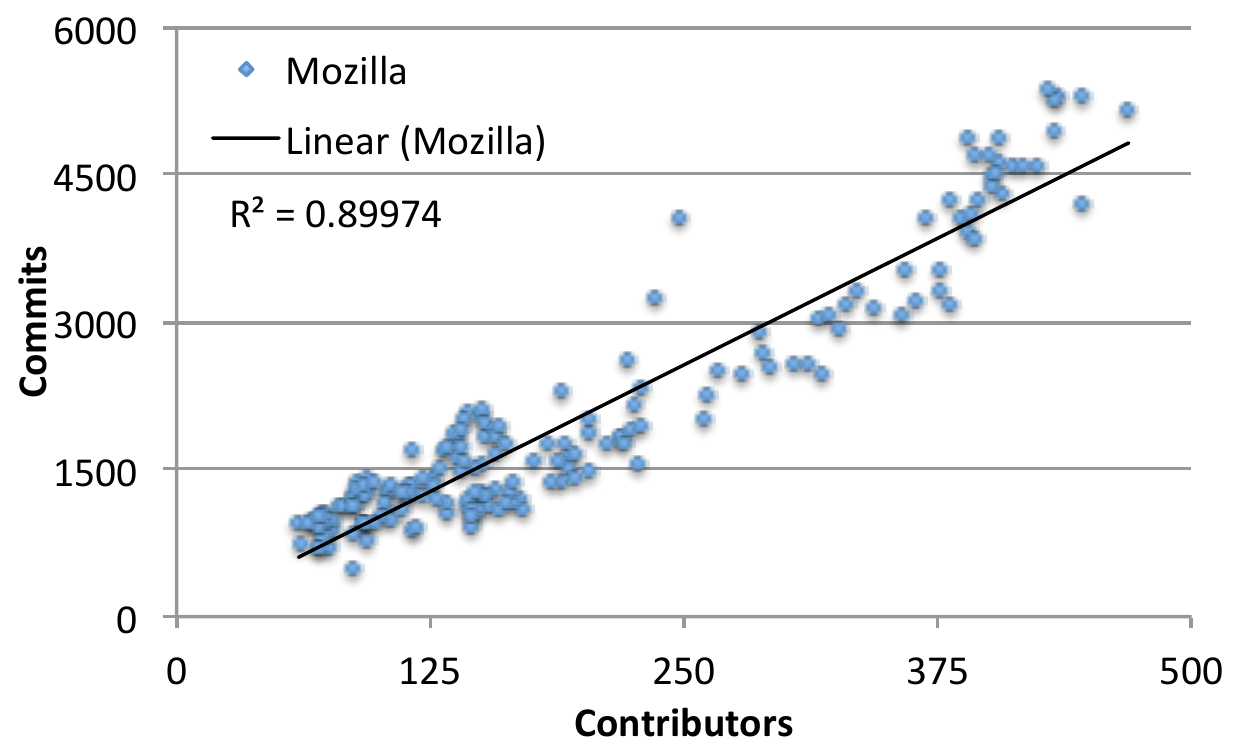}} 
  \subfloat[Puppet (\(\rho\)=0.69).]{ 
    \label{fig:subfig:spearman_puppet} 
    \includegraphics*[width=6cm]{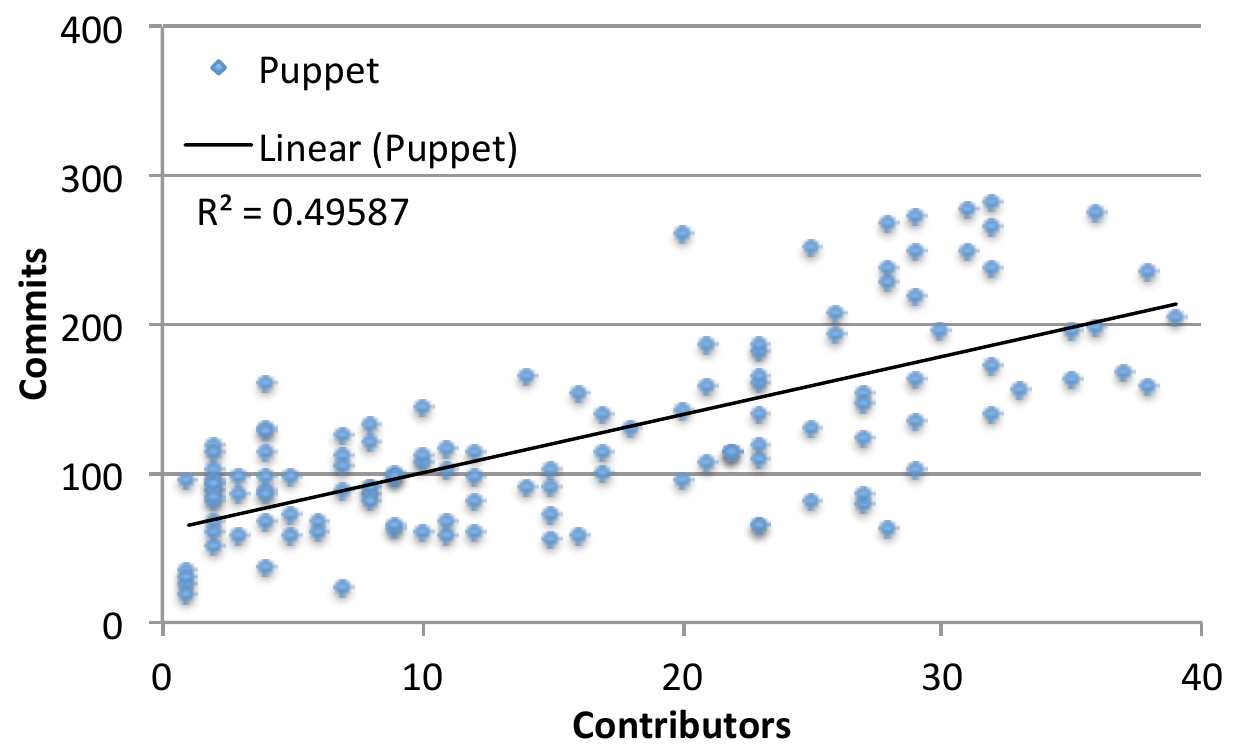}} 
 \vspace{1pt}
    \subfloat[Eucalyptus (\(\rho\)=0.54).]{ 
    \label{fig:subfig:spearman_eucalyptus} 
    \includegraphics*[width=6cm]{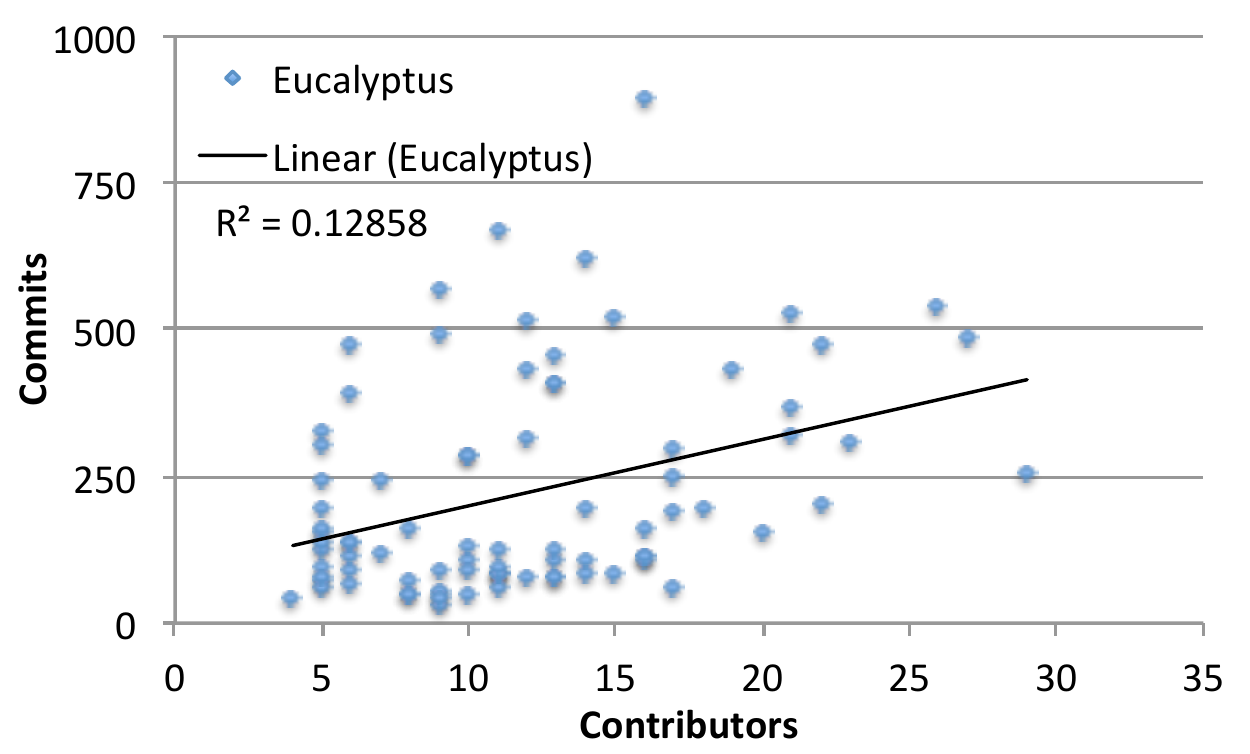}} 
  \subfloat[PHP (\(\rho\)=0.34).]{ 
    \label{fig:subfig:spearman_php} 
    \includegraphics*[width=6cm]{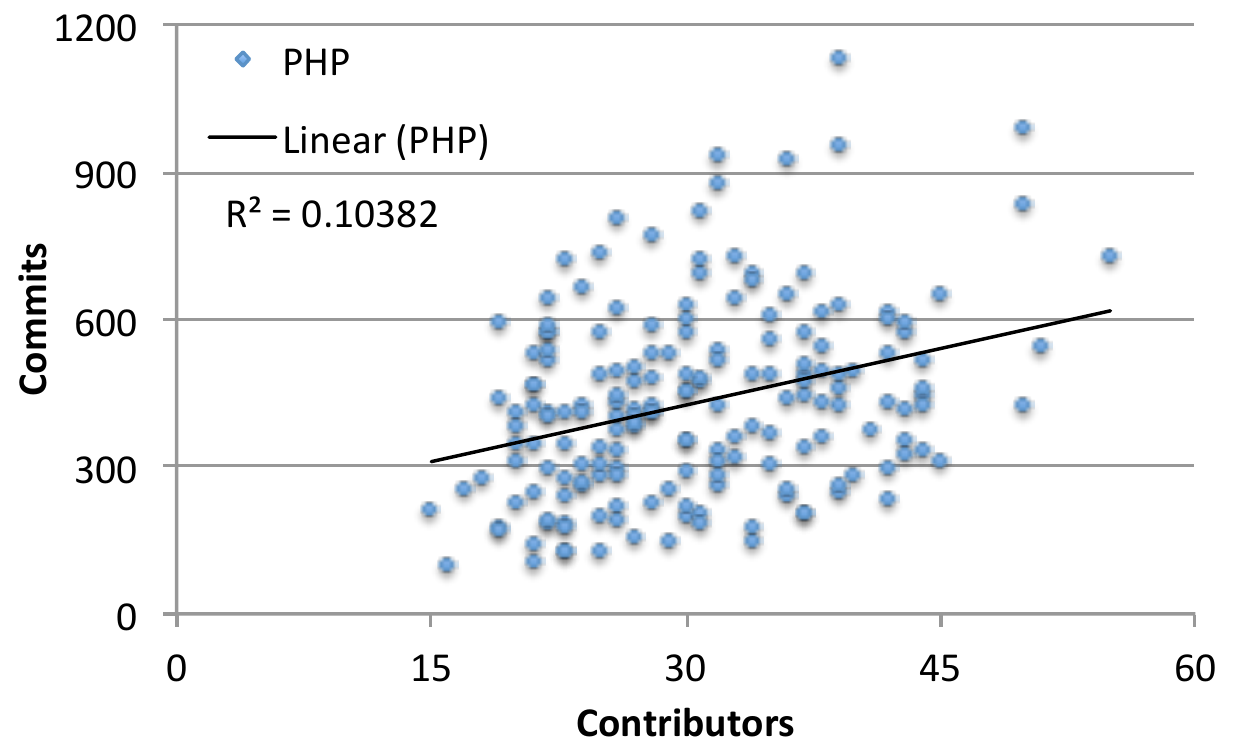}} 
  \subfloat[CloudStack (\(\rho\)=0.04).]{ 
    \label{fig:subfig:spearman_cloudstack} 
    \includegraphics*[width=6cm]{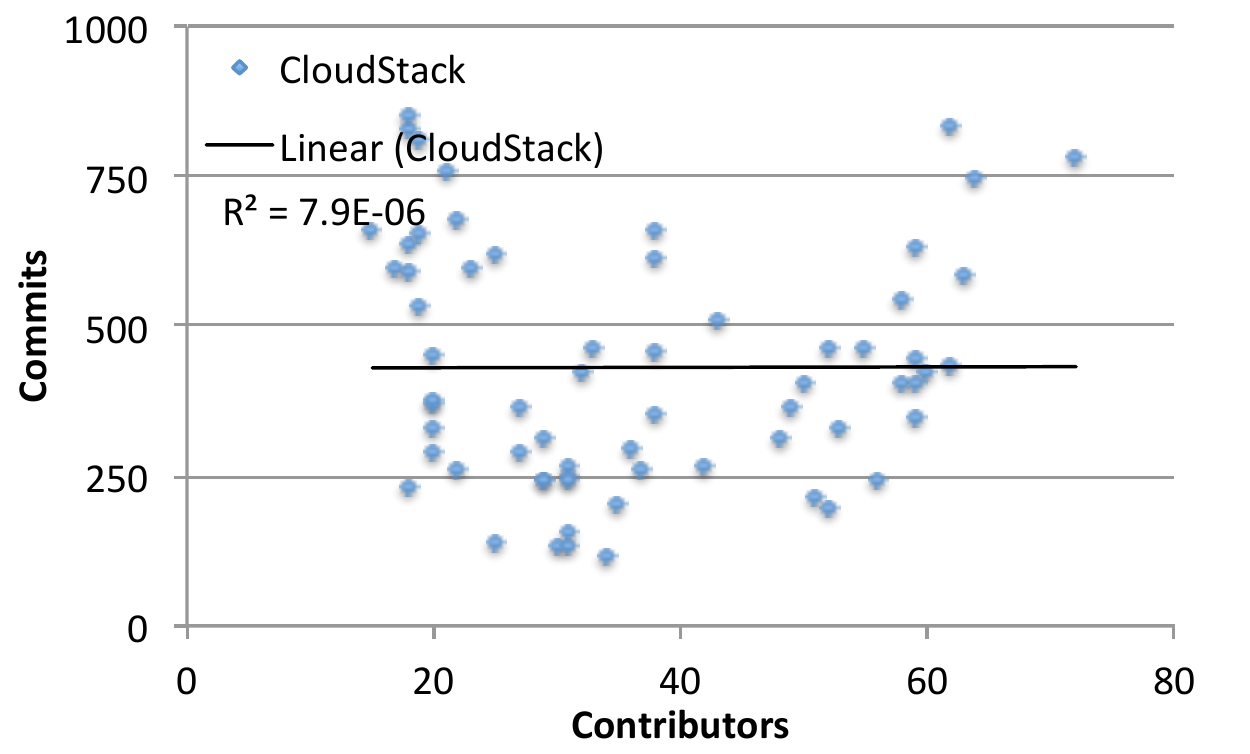}} 
\vspace{1pt}
  \caption{Productivity with respect to the number of active contributors and the number of commit activities.} 
  \label{fig:productivity_examples} 
\end{figure*}

Figure \ref{fig:productivity} shows the relationship between the size of the developer communities and their Spearman's rank correlation coefficient (productivity), for all 20 open source projects (Table \ref{tbl:population}). In general, the Spearman's rank correlation coefficient grows when the size of the developer community grows. This suggests for larger developer communities more active contributors usually brings more commit activities. The actual Spearman's coefficient values are presented in Table \ref{tbl:population}.

Further, to understand the implication of the Spearman's rank correlation coefficient, we plot the relationship between the number of active contributors and the number of commit activities for Gnome (\(\rho\)=0.96), Mozilla (\(\rho\)=0.77), Puppet (\(\rho\)=0.69), Eucalyptus (\(\rho\)=0.54), PHP (\(\rho\)=0.34), and CloudStack (\(\rho\)=0.04) in Figure \ref{fig:productivity_examples}. When the Spearman's rank correlation coefficient is greater, the relationship between the number of active contributors and the number of commit activities is stronger. As shown in the sub-figures for Gnome, Mozilla and Puppet, when the size of the developer community increases, the productivity of the developer community also increases in a predictable way. When the Spearman's rank correlation coefficient decreases, the relationship between the number of active contributors and the number of commit activities becomes weaker. As shown in the sub-figure for Eucalyptus, when the size of the developer community increases, the productivity of the developer community does not necessary increase. When the Spearman's rank correlation coefficient further decreases, the relationship between the number of active contributors and the number of commit activities becomes non-deterministic. As shown in the sub-figures for PHP and CloudStack, the data presented seems to be randomly distributed. 

\begin{figure*}[t!]
\centering
  \subfloat[Community Size and Diversity Index.]{ 
    \label{fig:subfig:dp-a} 
    \includegraphics*[width=8.3cm]{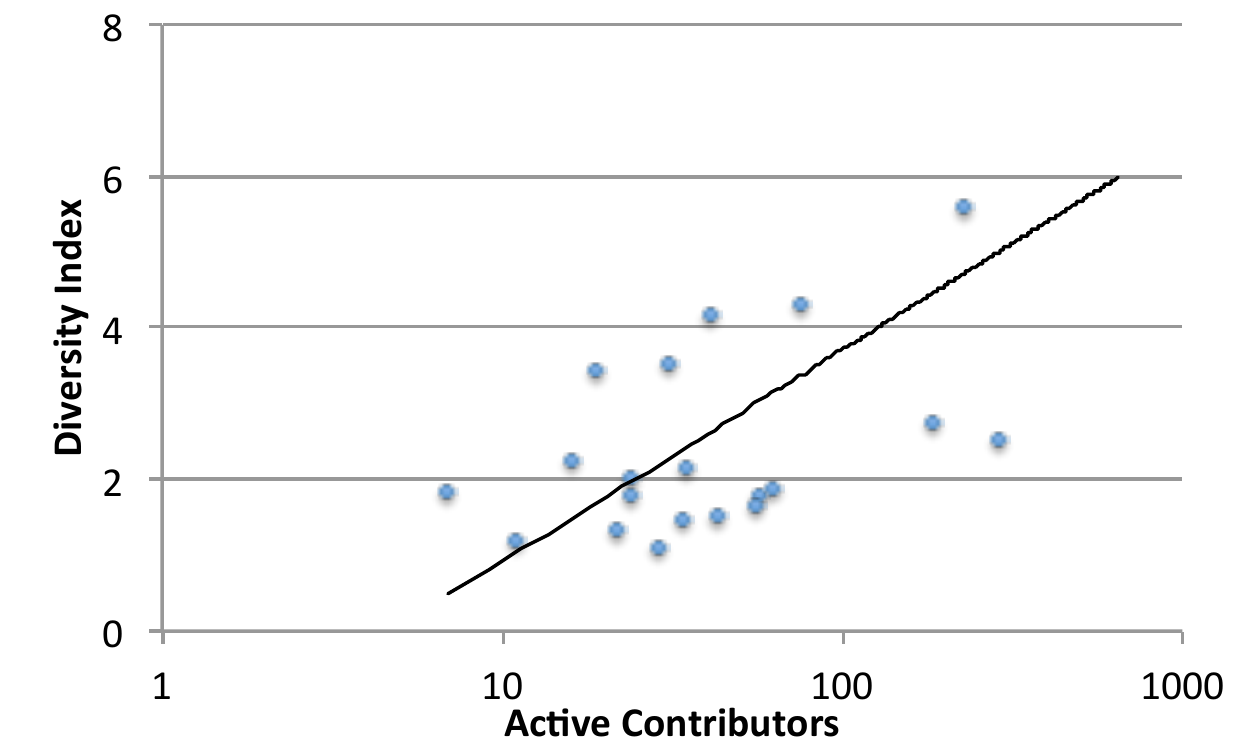}} 
 \subfloat[The Relationship between Diversity and Productivity.]{ 
      \includegraphics[width=8.3cm]{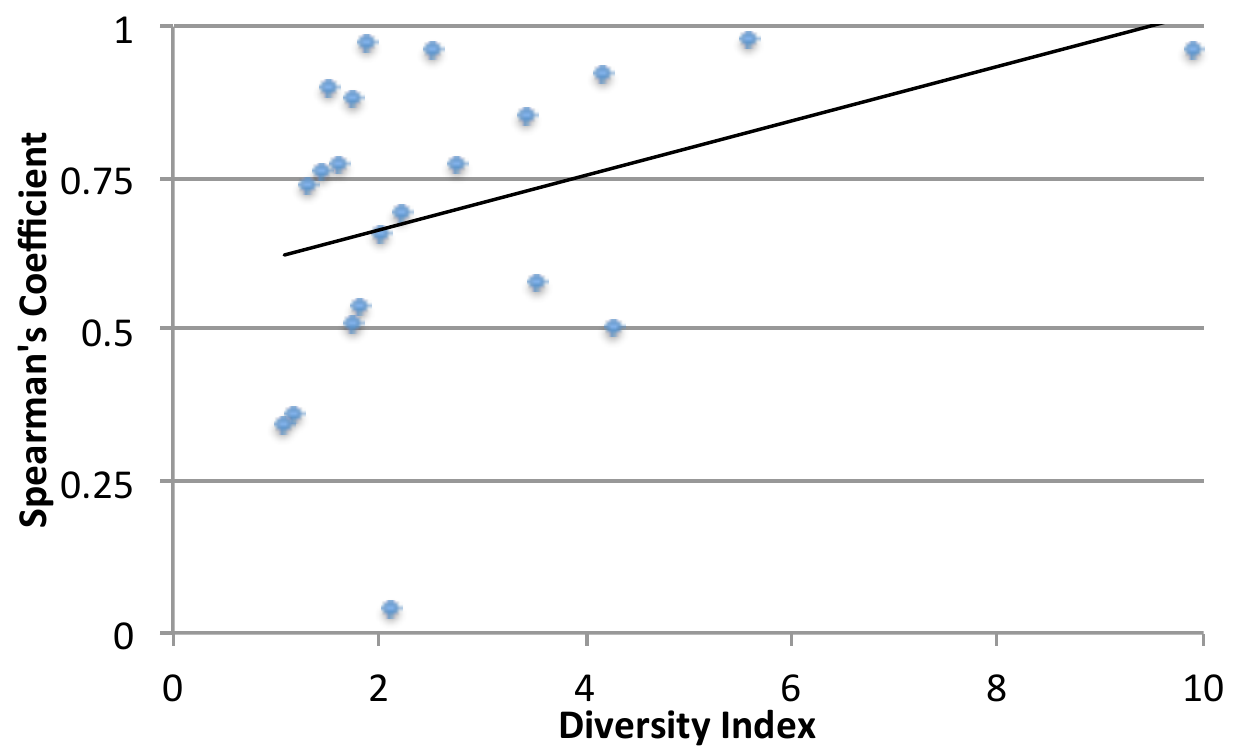}
  \label{fig:subfig:dp-b}} 
  \caption{Diversity of Open Source Developer Communities.} 
  \label{fig:diversity_productivity} 
\end{figure*}

For each sub-figure, we also plot a linear regression trend line, along with the R-squared value of the trend line. R-squared is a statistical measure of how close the data are to the fitted regression line. It is also known as the coefficient of determination, or the coefficient of multiple determination for multiple regression. However, the R-squared value does not effectively describe the skewness of the data distribution. For example, the Mozilla project has a higher R-squared value (0.89974) then the Gnome project (0.81462), but a visual inspection suggests that the data points for the Gnome project distributes more evenly on both sides of the linear regression trend line. Therefore, the Spearman's rank correlation coefficient is a better parameter to describe the relationship between the number of active contributors and the number of commit activities. 

\section{Diversity}
\label{sec:diversity}

Many open source projects have a dominating organization contributing the majority of the commit activities, while other contributing organizations are minor players. We define the diversity index of an open source developer community as square root of the inverse Simpson's index \cite{simpson1949measurement}. The diversity index can be calculated using the following equations:
\begin{equation}
\label{eq:simpson}
S=\sum_{i=0}^{i=n} p_{i}^{2},
\end{equation}
\begin{equation}
\label{eq:diversity}
D=\sqrt{\frac{1}{S}},
\end{equation}
\\where \textit{S} is the Simpson's index and \textit{D} is the diversity index, \textit{n} is the total number of contributing organizations in the community, while \textit{p\textsubscript{i}} is the contribution from the \textit{i\textsuperscript{th}} organization. The greater the diversity index, the greater is the diversity of the open source developer community.

Figure \ref{fig:subfig:dp-a} shows the relationship between the diversity index and the size of the developer community. In general, larger communities are more diverse than smaller communities. This is not a surprising observation---larger communities usually have more contributing organizations, leading to the increase in diversity. 

Figure \ref{fig:subfig:dp-b} shows the relationship between diversity and the Spearman's rank correlation coefficient (productivity). In particular, for projects with low diversity (such as Eucalyptus and MySQL with their diversity index values of 1.83 and 1.63, respectively as shown in Table \ref{tbl:population}), the development process is tightly controlled by the primary contributing organization. Depending on the way the primary contributing organization manages its software development team, the productivity of the software development team can vary significantly. This is reflected by the wide range of the Spearman's rank correlation coefficient for these projects. In many cases, the primary contributing organization assigns software development tasks to its internal team members according to its internal development plan. As a result, a particular developer's contribution to the project is limited by his/her job assignment. In a sense, this prevents a particular developer from contributing more code to the project, because certain areas or features that he/she is capable of working on happen to be out of the scope of his/her job assignment. 

\begin{figure*}[t!]
\centering
 \subfloat[Normal Growth.]{ 
    \label{fig:subfig:1a}  
    \includegraphics*[width=6cm, viewport=5 0 400 250]{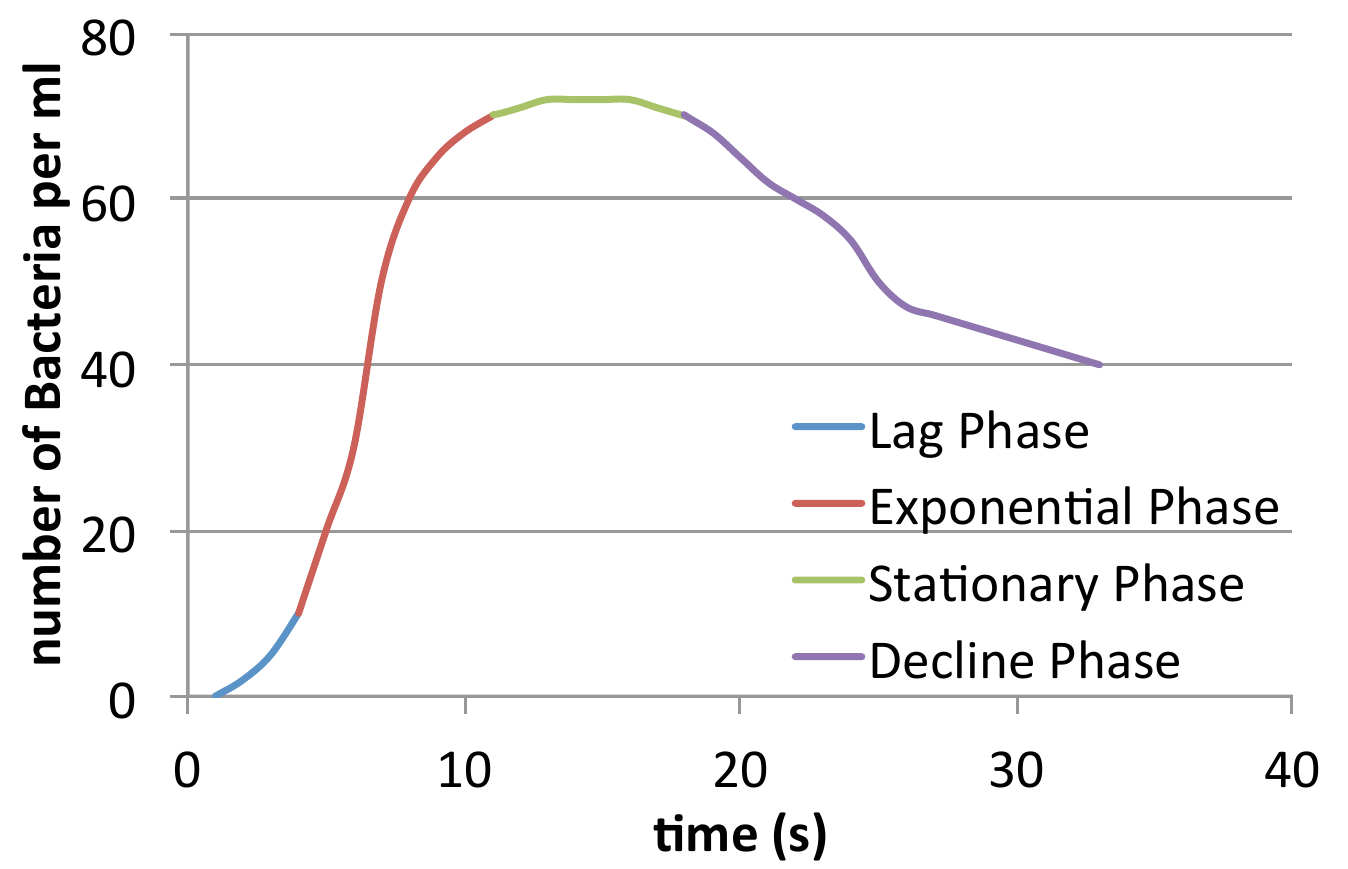}} 
    \hspace{5pt}
  \subfloat[Bi-Phase Growths.]{ 
    \label{fig:subfig:1b} 
    \includegraphics*[width=5.7cm, viewport=22 0 400 250]{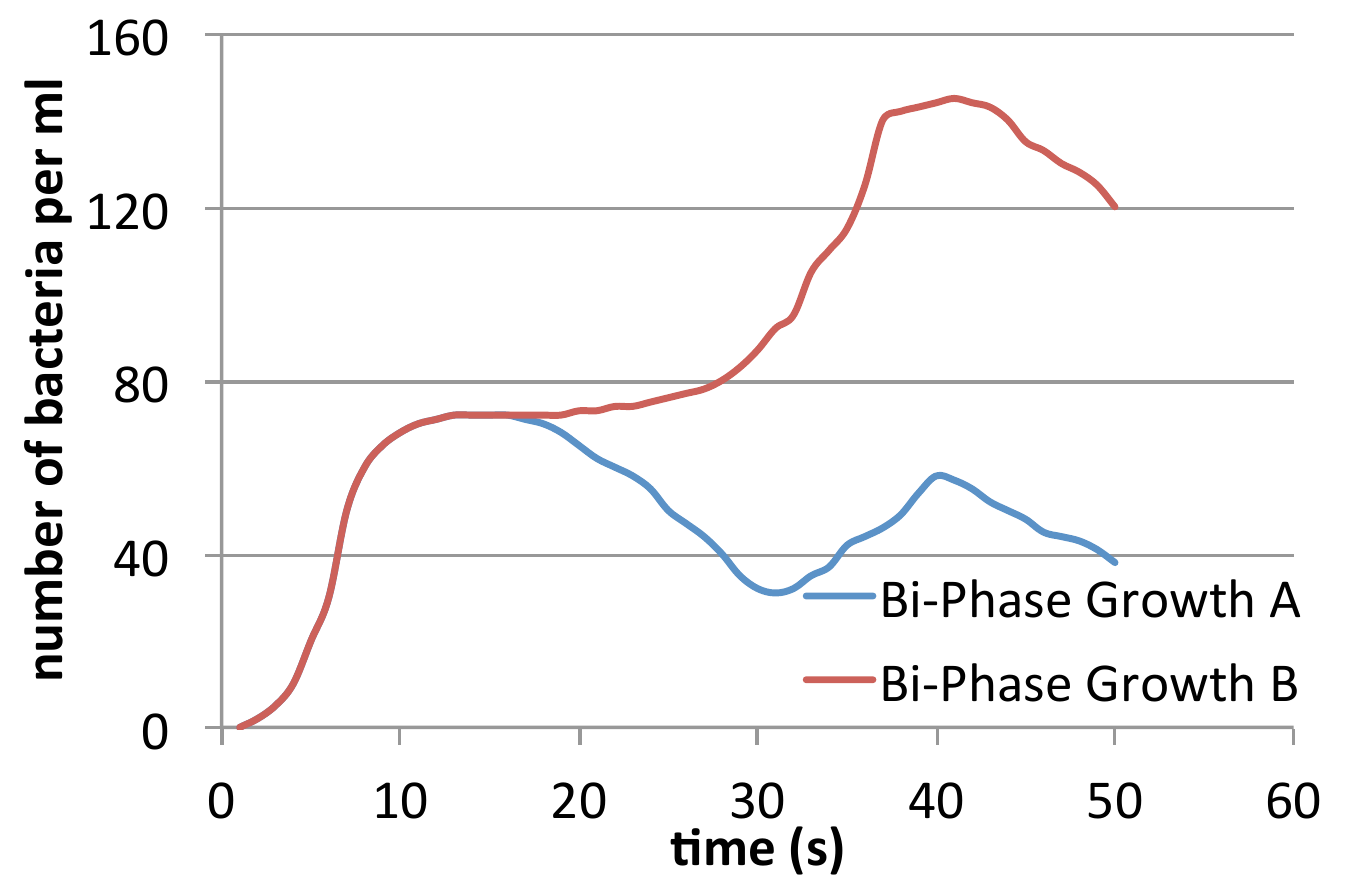}} 
        \hspace{5pt}
  \subfloat[Growth Curves.]{ 
    \label{fig:subfig:1c} 
    \includegraphics*[width=5.7cm, viewport=22 0 400 250]{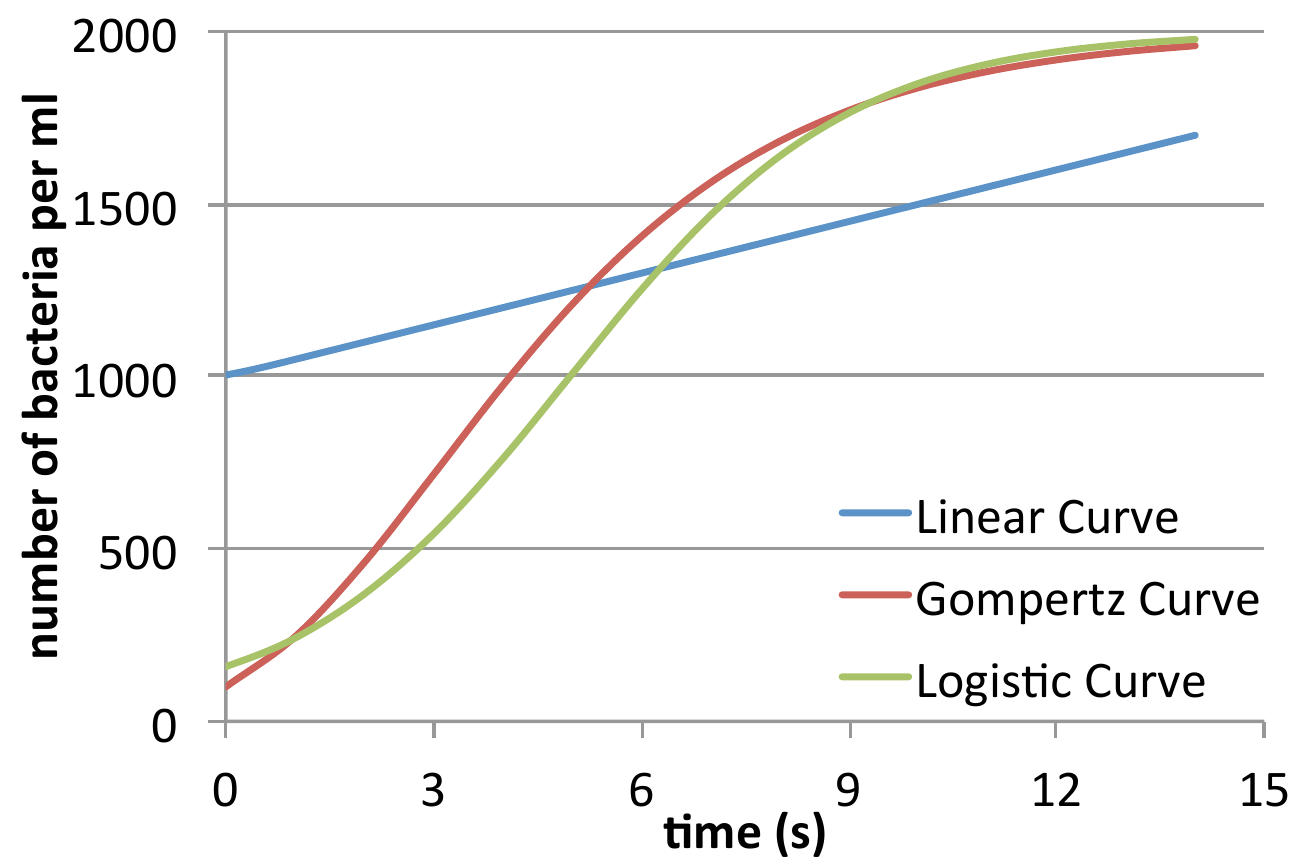}} 
    \caption{Examples of Bacteria Growth Curves.} 
    \vspace{-5pt}
  \label{fig:growth_bacteria} 
\end{figure*}

For projects with medium diversity (such as Apache Spark and Mozilla of 4.18 and 2.76, respectively), there usually exists a primary contributing organization contributing around 50\%  of the commit activities, with one or more secondary contributing organizations each contributing around 10\% to 30\% of the commit activities. In an open source developer community with multiple significant contributing organizations, leadership is usually defined by the code contributions (``show me the code" \cite{raymond1999cathedral} \cite{raymond2001cathedral}). To maintain or compete for leadership in the developer community, both the primary and secondary contributing organizations are motivated to contribute more code to the project. Since the development plan is no longer controlled by the primary contributing organization, a particular developer enjoys more freedom in that he/she can contribute code to multiple areas or features in which he/she is capable of. This is reflected in the relatively high value of the Spearman's rank correlation coefficient - growth in the size of the developer community positively increase the productivity of the open source project. 

For projects with high diversity (such as OpenStack and the Linux kernel of 5.60 and 9.90, respectively), there no longer exists a primary contributing organization with a dominating contribution. Instead, there exists multiple major contributing organizations with approximately the same amount of code contributions. More importantly, the sum of the contributions from these major contributing organizations does not represent a dominating force among the developer community. For example, during the first quarter of calendar year 2015, the top 5 contributing  organizations for OpenStack was RedHat (7.3\%), IBM (5.0\%), Mirantis (4.7\%), HP (4.6\%), and Rackspace (1.6\%) \cite{jiang2015q1iaas}. The sum of the contributions from the top 5 contributing organizations was only 23.2\%. In this case, smaller contributing organizations develop a feeling of being equally important in the developer community. They are motivated to contribute more code to increase their visibility in the developer community. At the same time, software development tasks are no longer "owned by" or "assigned to" a particular group of developers. This means a particular developer has more liberty in working on multiple areas or features, allowing him/her to contribute more code to the project. For a particular business, it is expected to have multiple highs and lows during its various phases. For open source projects with low or medium diversity, a small number of contributing organizations dominate the development process. The highs and lows of these contributing organizations have a significant impact on the productivity of the developer community. For open source projects with high diversity, there exists no dominating contributing organizations in the developer community. As a result, the highs and lows of a single contributing organization does not have a significant impact on the productivity of the developer community. This is reflected in the higher value of the Spearman's rank correlation coefficient - growth in the size of the developer community always increase the productivity of the open source project. 

\section{Growth}
\label{sec:growth}

Growth curves have been applied to a wide range of disciplines such as biology and medical science. Most organic matters grow with successive lag, exponential, stationary, and decline phases. A good example of such patterns is the growth of bacteria in a confined nutrient-containing broth, as shown in Figure \ref{fig:subfig:1a}. In the lag phase, bacteria become mature but are not yet able to reproduce. The exponential phase is characterized by cell doubling, where the number of new bacteria appearing per unit time is proportional to the present population. In the stationary phase, the growth rate is equal to the death rate, which is usually the result of nutrient depletion or spatial confinements. In the decline phase, bacteria run out of nutrient, and die. Bacteria exhibit multiple exponential growth phases under certain circumstances, which are called bi-phase growth or multi-phase growth (Figure \ref{fig:subfig:1b}). 

A number of mathematical models have been proposed to represent such growth. Among them, the Gompertz curve \cite{winsor1932gompertz} and the logistic curve \cite{volterra1938population} -- as shown in Figure \ref{fig:subfig:1c} -- have been adopted in a variety of disciplines. The formula for the Gompertz curve and the logistic curve are presented in Equations \ref{eq:eqa} and \ref{eq:eqb} respectively, where \textit{y*} represents the upper limit of the growth curves, while $\alpha$ is a parameter closely related to the growth rate. 
\begin{equation}
\label{eq:eqa}
\frac{dy}{dx} = \alpha y (ln y^{*} - ln y).
\end{equation}
\begin{equation}
\label{eq:eqb}
\frac{dy}{dx} = \alpha y (y^{*} - y).
\end{equation}

Both the Gompertz curve and the logistic curve work well in describing the lag, exponential, and stationary phases of organic growths. They are not applicable when the growth has entered the decline phase, nor are they capable of describing bi-phase or multi-phase growths. Examples of applications include bacteria growth in biology \cite{zwietering1990modeling}, tumor growth in medical science \cite{laird1964dynamics}, population growth in social science \cite{montroll1978social}, and market growth in economics \cite{meade1984use} \cite{mahajan1990new}.

The growth of open source developer communities is similar to the growth of bacteria in a confined nutrient-containing broth in that: (a) the speed of technology diffusion is related to the current size of the developer community \cite{bonaccorsi2003open} \cite{o2007emergence}; (b) financial and human resources are needed to sustain the growth of the communities; and (c) there are growth limits for the communities due to technological, economical, social, and resource constrains. Therefore, growth curves established in other disciplines provide a solid ground on which we can study the growth of open source developer communities. 

\begin{figure*}[t!]
\centering
 \subfloat[Small Size Projects.]{ 
    \label{fig:subfig:gc1}  
    \includegraphics*[width=8.4cm]{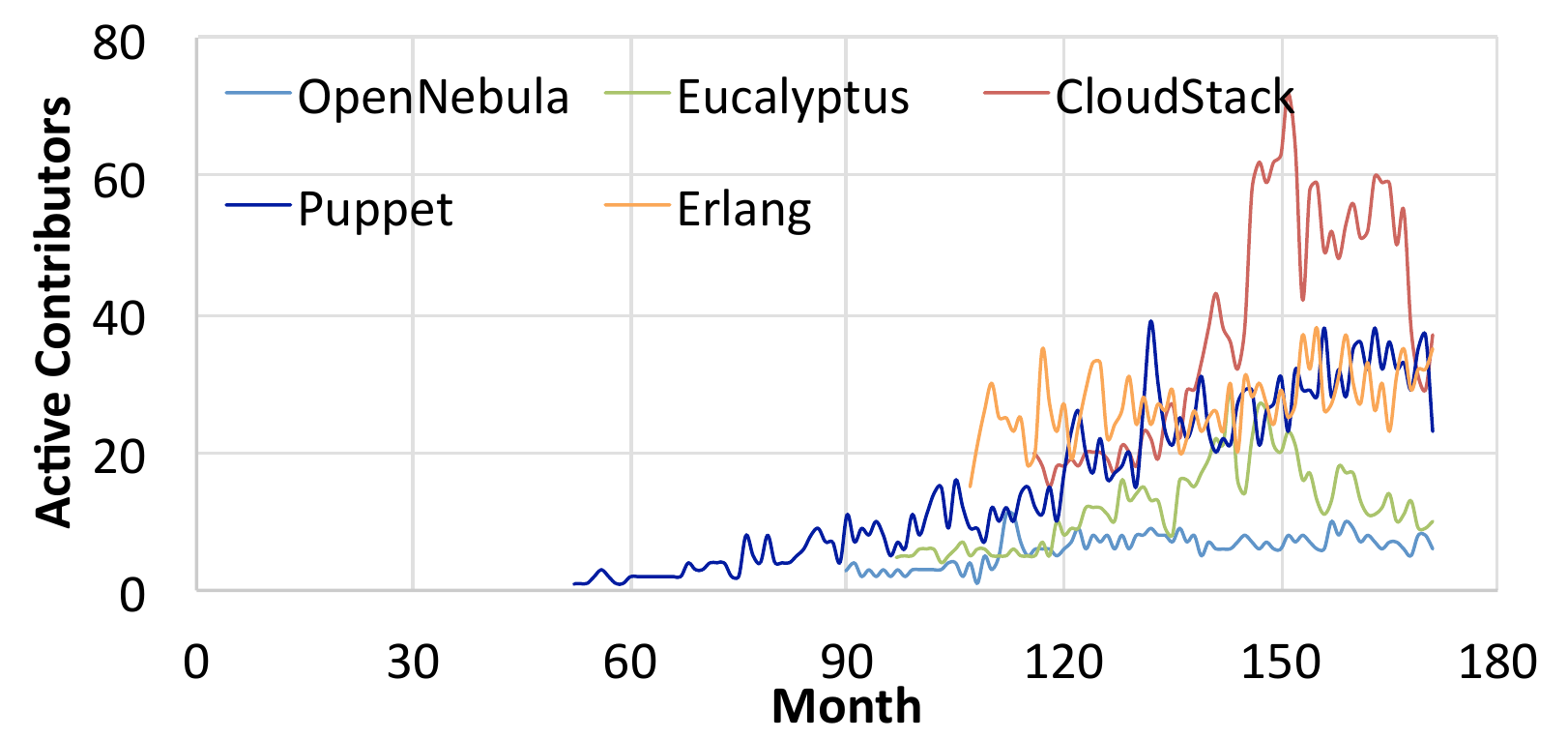}} 
 \subfloat[Small Size Projects.]{ 
    \label{fig:subfig:gc2}  
    \includegraphics*[width=8.4cm]{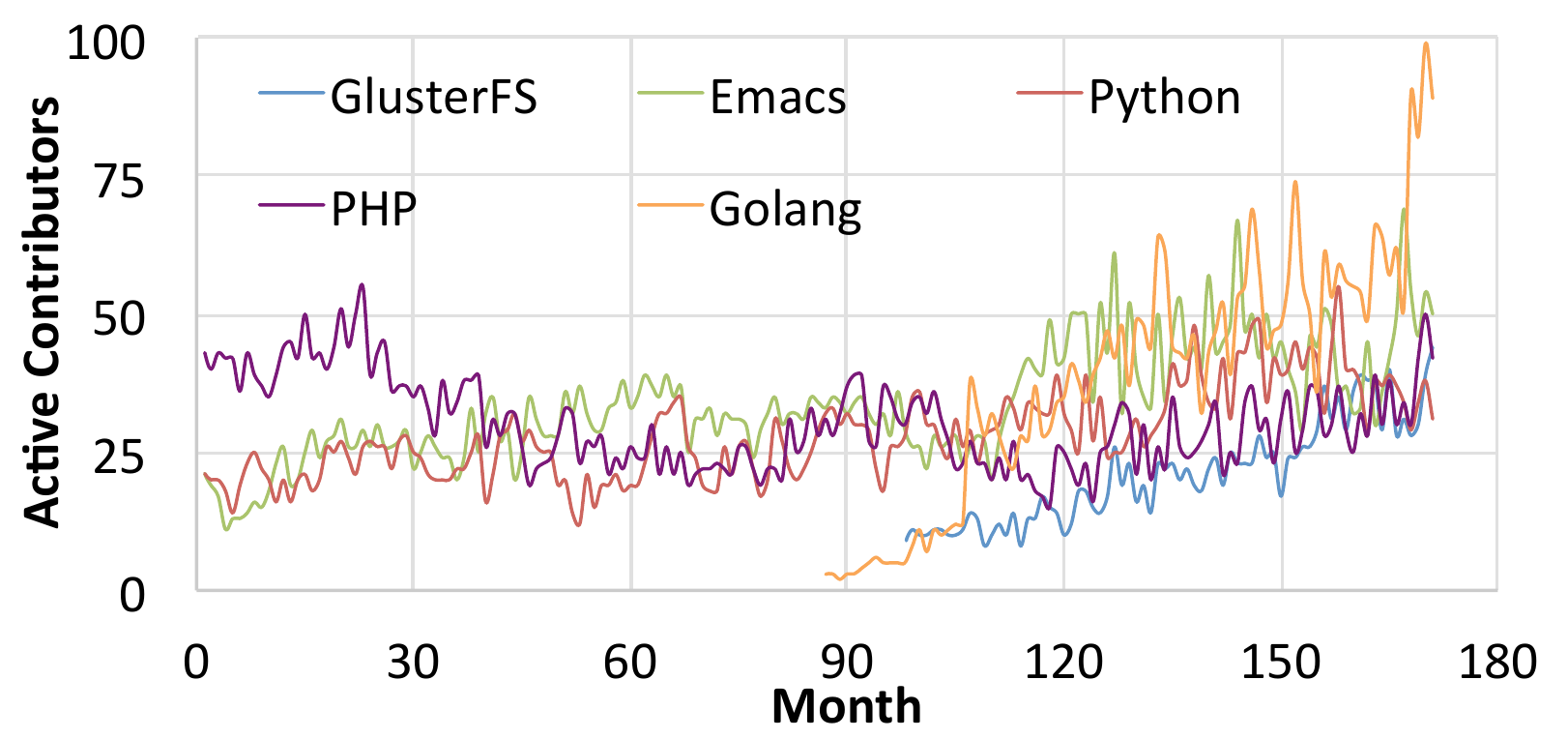}} 
    \vspace{10pt}
  \subfloat[Medium Size Projects.]{ 
    \label{fig:subfig:gc3} 
    \includegraphics*[width=8.4cm]{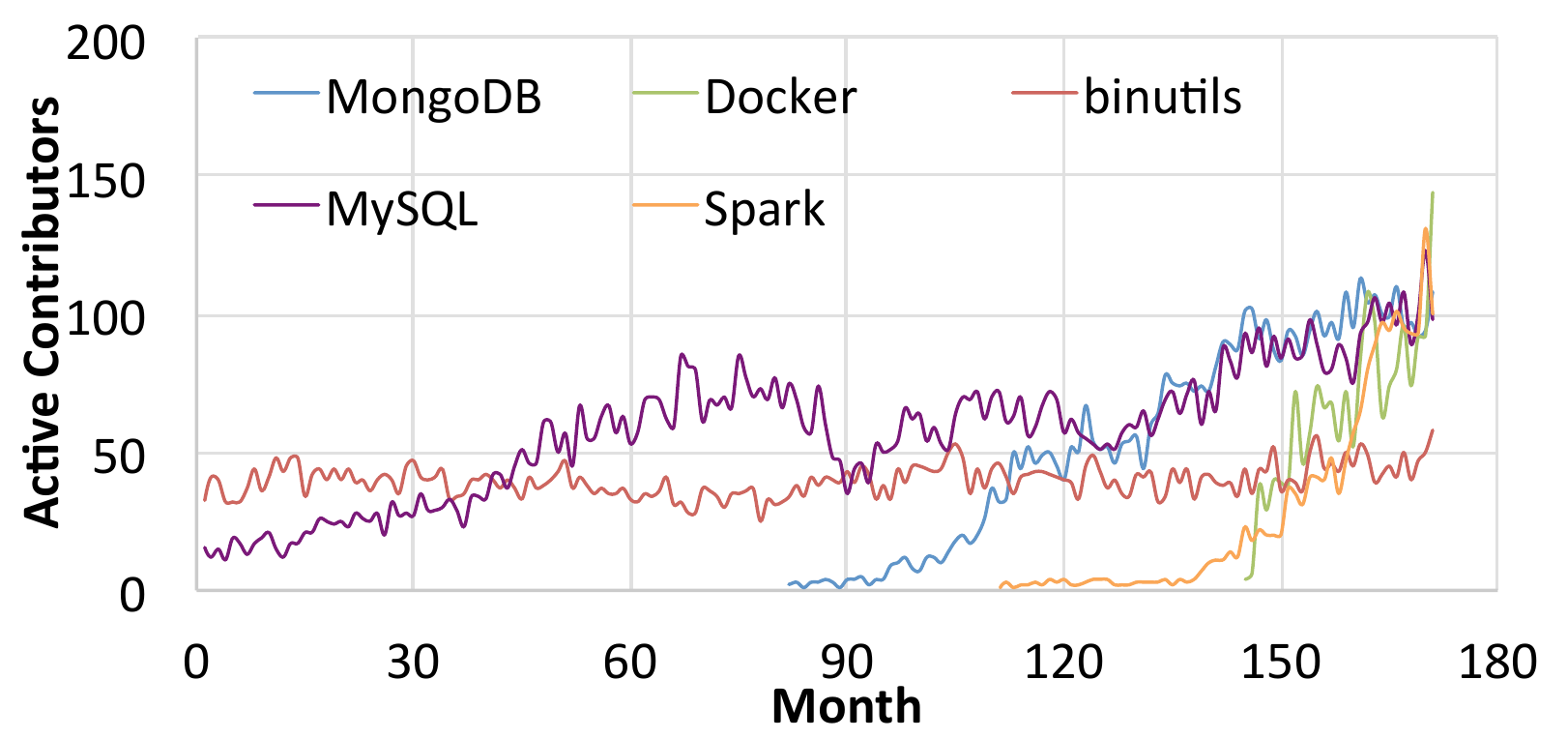}} 
  \subfloat[Medium Size Projects.]{ 
    \label{fig:subfig:gc41} 
    \includegraphics*[width=8.4cm]{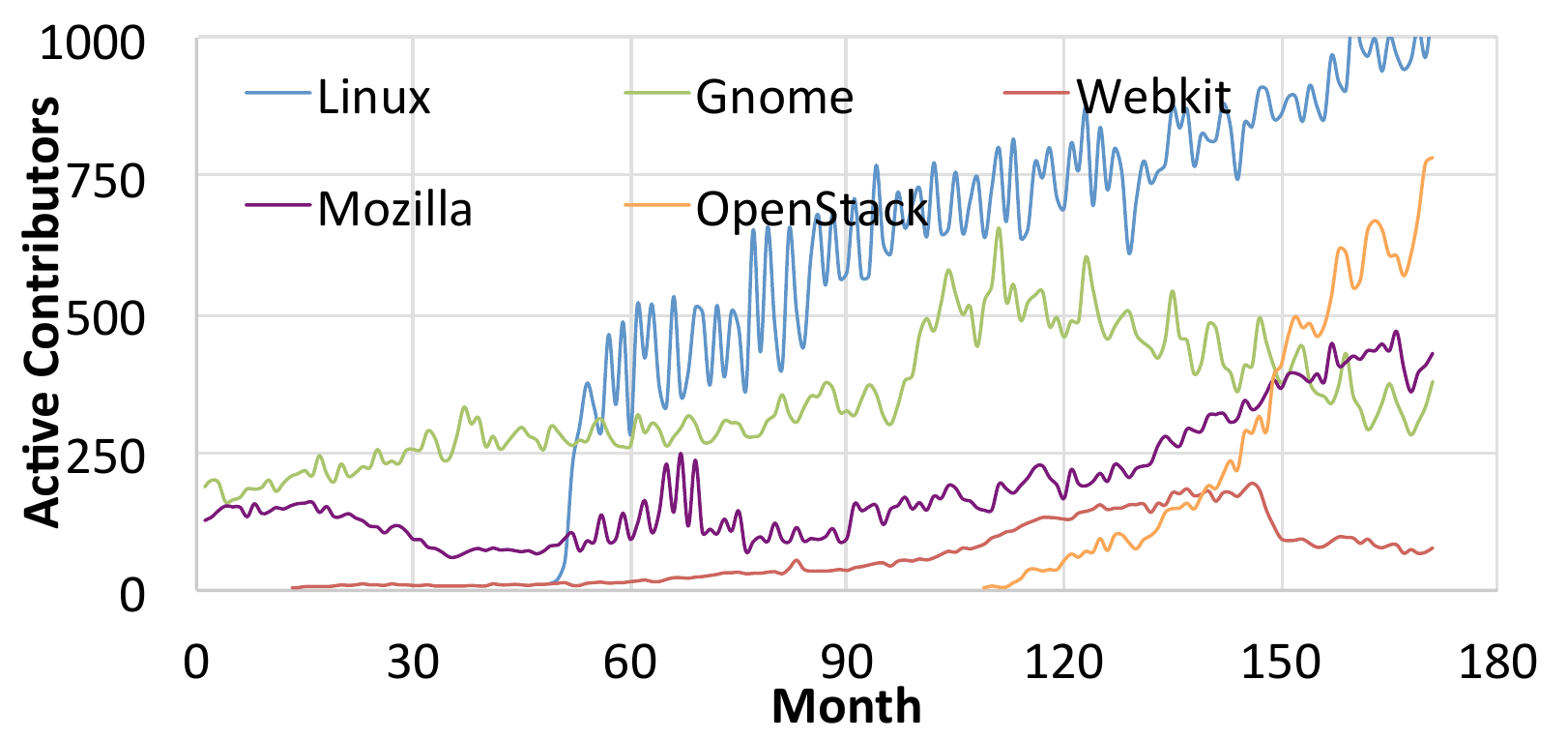}} 
    \vspace{5pt}
  \caption{Growth Curves of Open Source Projects with respect to Different Project Sizes.} 
  \label{fig:growth_curves} 
\end{figure*}

Figure \ref{fig:growth_curves} shows the growths in the number of active contributors for 20 selected open source projects. To make the figure easier to read, we break the 20 projects into four sub-figures according to their number of active contributors. The horizontal axis represents the number of months starting from January 2001. For example, 60 means December 2005 and 120 means December 2010. The vertical axis represents the number of active contributors in each month. In all sub-figures, we observe a mixture of different growth behaviors at different stages. For example, Linux, Spark and OpenStack seem to be in their exponential phase; Erlang and Puppet seem to be in their stationary phase; Eucalyptus and CloudStack seem to be in their decline phase; MySQL and PHP seem to exhibit characteristics of bi-phase growth behaviors. 

\begin{figure*}[t!]
\centering
 \subfloat[CloudStack.]{
    \label{fig:subfig:growth_dev_cloudstack}  
    \includegraphics*[width=8.3cm]{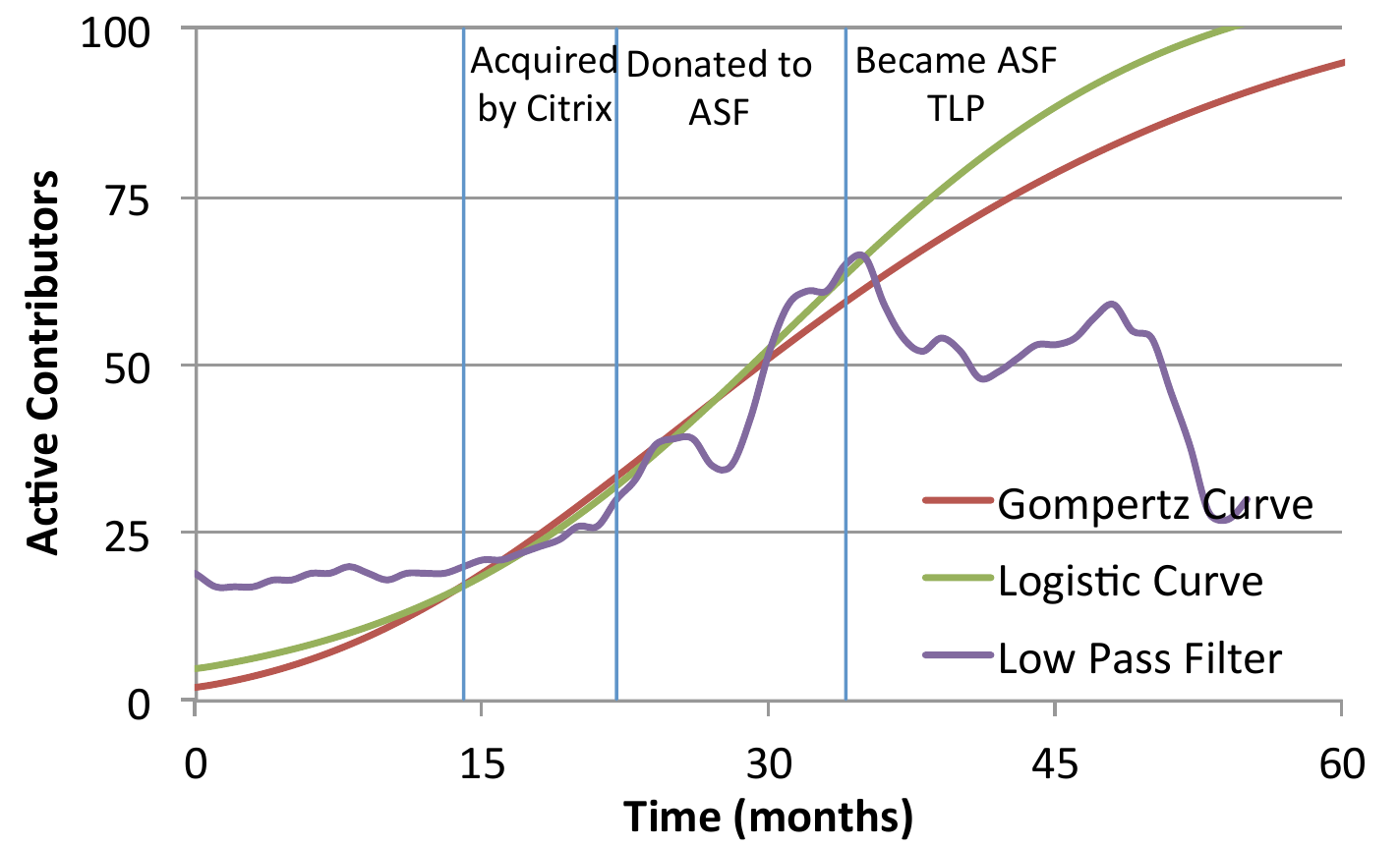}} 
 \subfloat[Eucalyptus.]{ 
    \label{fig:subfig:growth_dev_eucalyptus}  
    \includegraphics*[width=8.3cm]{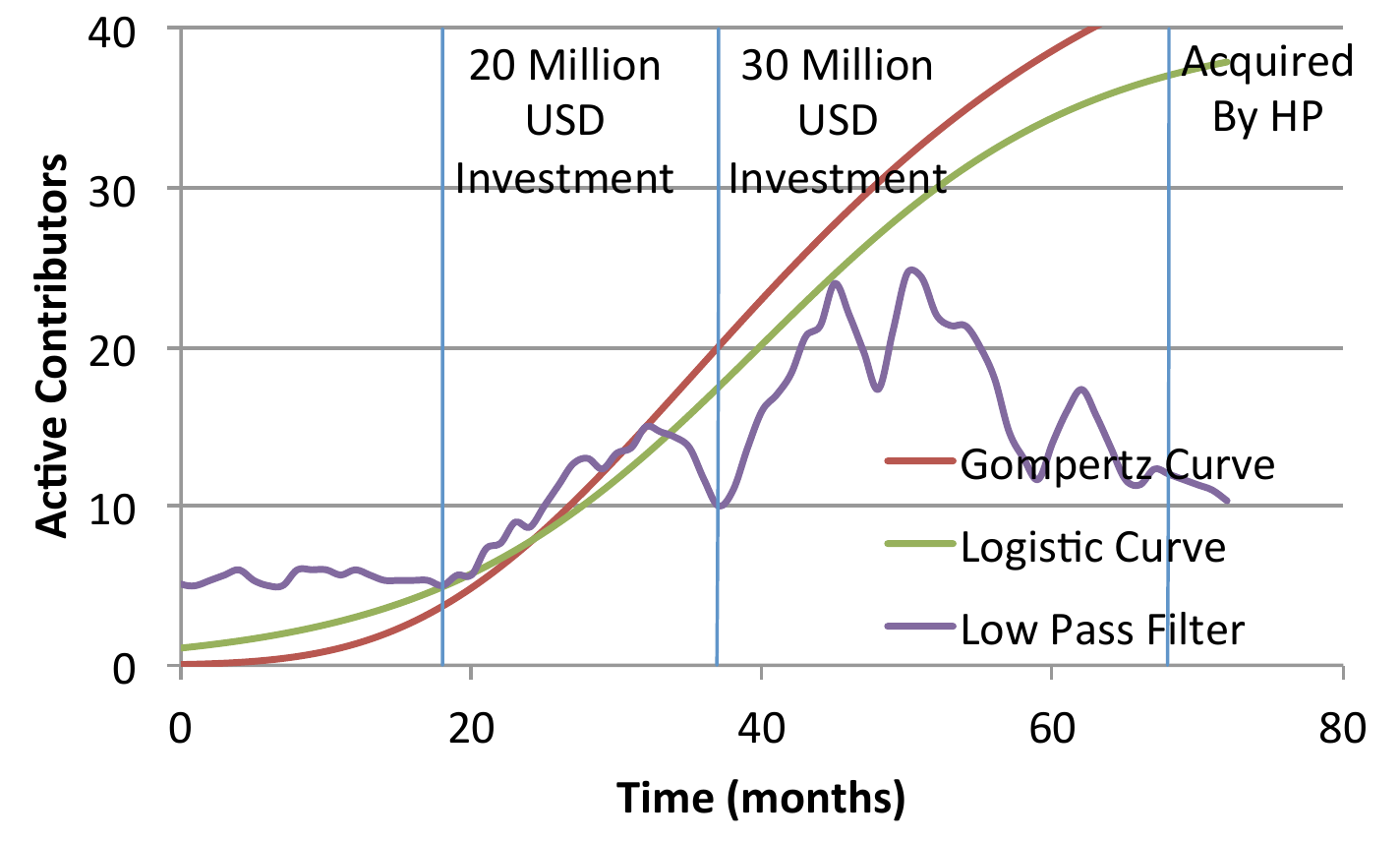}} 
    \vspace{5pt}
 \subfloat[OpenNebula.]{ 
    \label{fig:subfig:growth_dev_opennebula}  
    \includegraphics*[width=8.3cm]{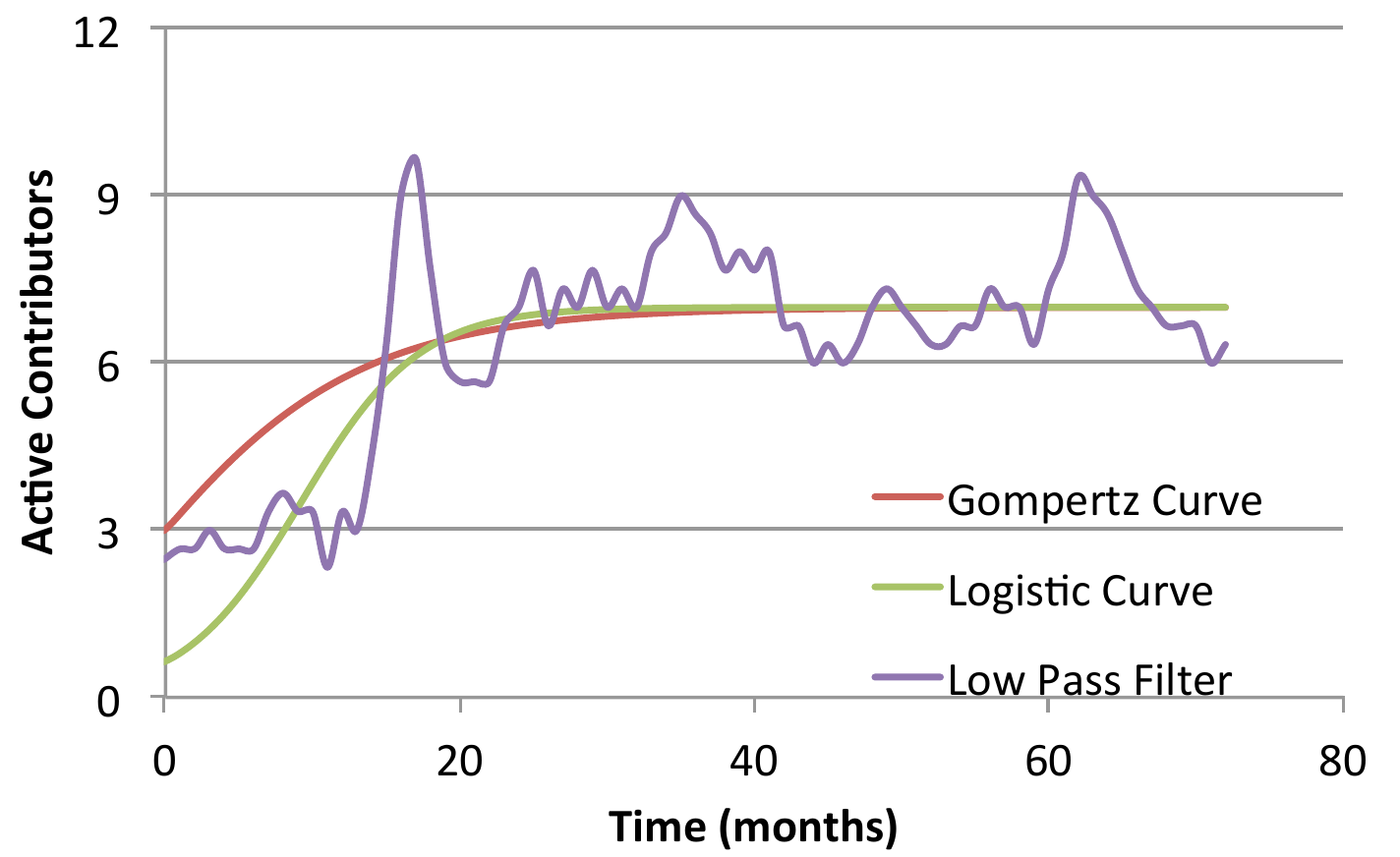}} 
 \subfloat[OpenStack.]{ 
    \label{fig:subfig:growth_dev_openstack}  
    \includegraphics*[width=8.3cm]{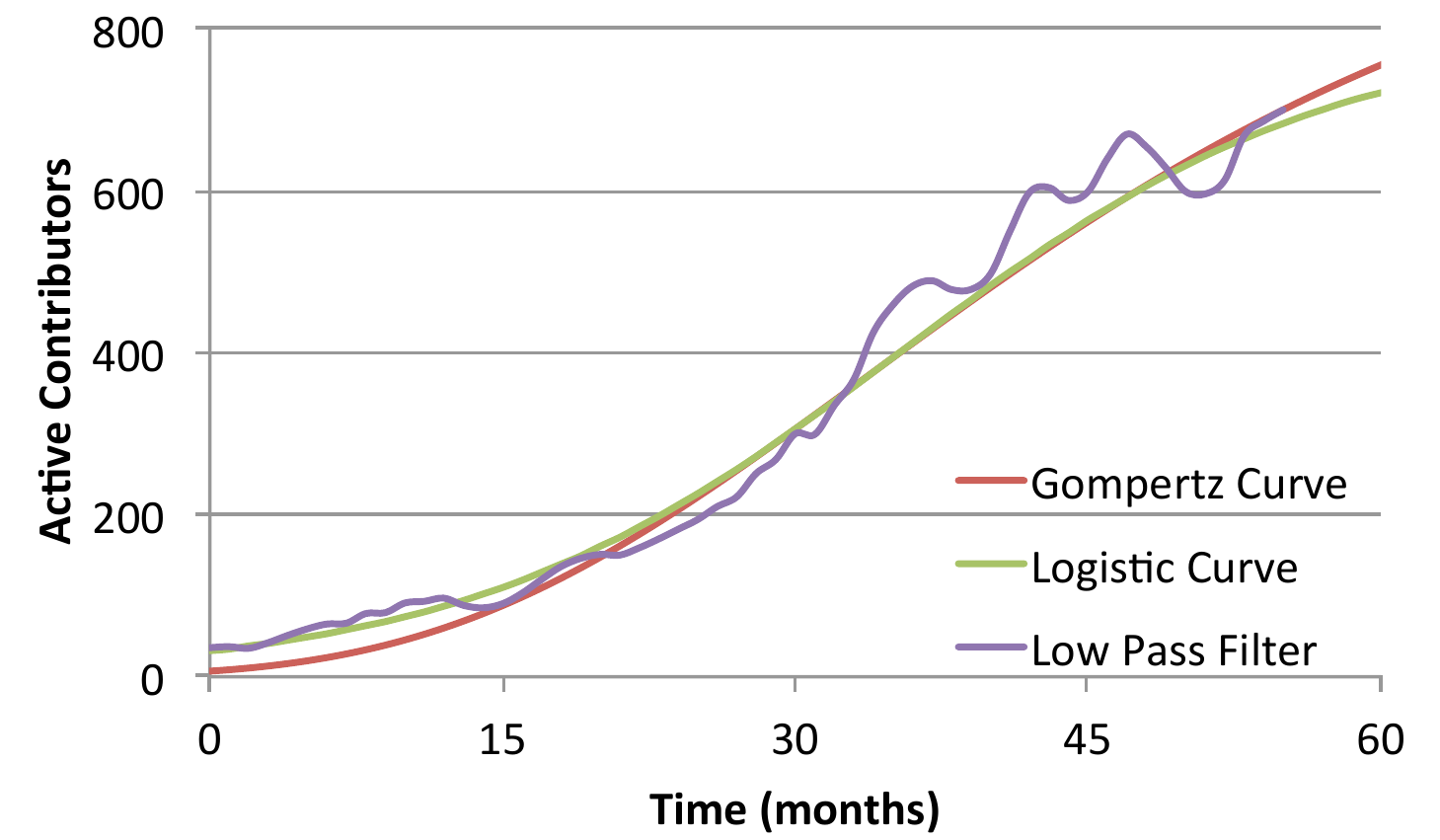}} 
  \caption{The Growth of Open Source IaaS Developer Communities.} 
  \label{fig:growth_developer} 
\end{figure*}

\textbf{Case Studies.} To better understand the relationship between the above-mentioned metrics, we use four major open source projects in the ares of Infrastructure-as-a-Service (IaaS) as case studies. These projects are CloudStack, Eucalyptus, OpenNebula, and OpenStack. Some of the events -- such as acquisitions, donations, capital investments -- mentioned in this case study come from Jiang's quarterly community analysis of these four open source projects, which was a 4-year effort started in 2012 \cite{jiang2014open}. Figure \ref{fig:growth_developer} shows the growths in the number of active contributors for these four projects. It should be noted that t(0) in these figures represents the month in which the project was launched, therefore t(0) is different for different projects. To account for the fluctuation in the number of active contributors from month to month, we use a low pass filter to pre-process the data. The low pass filter is a moving average window filter, with the window size set to 3 months. Based on the filtered data, we derive the key parameters of the growth curves for the developer communities, and plot them on the same figure along with the filtered data. 

The \emph{CloudStack} project has a total number of 370 developers, with 20 to 70 active contributors and 200 to 800 commit activities in each month. The community has a low diversity index of 2.14, and the Spearman's rank correlation coefficient between the number of active contributors and the number of commit activities is 0.04. The growth of the CloudStack developer community (Figure \ref{fig:subfig:growth_dev_cloudstack}) exhibits characteristics of a bi-phase growth. Before being acquired by Citrix, CloudStack was developed by a core development team, which is reflected in the flat line during the lag phase. After being donated to ASF, the project gained increasing level of attention from developers outside of Citrix, which is reflected in the exponential phase. However, the project started to lose momentum and enter its decline phase after becoming a top level project. Despite a minor revamp around the third quarter in 2014, the size of the developer community is quickly shrinking. Unless new resource is being invested into the project, it is expected that the project will continue to wither away. 

The \emph{Eucalyptus} project has a total number of 80 developers, with 10 to 40 active contributors and 100 to 400 commit activities in each month. The community has a low diversity index of 1.83, and the Spearman's rank correlation coefficient between the number of active contributors and the number of commit activities is 0.54. The growth of the Eucalyptus developer community (Figure \ref{fig:subfig:growth_dev_eucalyptus}) also exhibits characteristics of a bi-phase growth. Started as a university research project, Eucalyptus also had a stable development team before its first release, which is reflected by the flat line during the lag phase. Later on the community exhibits multiple grows and falls. A closer look into the company's funding history reveals that the starting point of both exponential phases coincide with the time when the company received investments from venture capitals. Similar to the growth of bacteria in a nutrient-containing broth, the community grows when nutrition (funding) is abundant and declines when nutrition depletes. A third growth did not occur when HP acquired Eucalyptus, suggesting that the acquisition did not inject new resource to revive the project. 

The \emph{OpenNebula} project has a total number of 80 developers, with 5 to 10 active contributors and 50 to 100 commit activities in each month. The community has a low diversity index of 1.11, and the Spearman's rank correlation coefficient between the number of active contributors and the number of commit activities is 0.64. The OpenNebula developer community (Figure \ref{fig:subfig:growth_dev_opennebula}) only has 10 active contributors at its maximum, which makes regression analysis difficult. The project maintains a stable size of 7 active contributors most of the time. This seems to be the size of the OpenNebula development team at C12G Labs, which does not change much over time. In other words, the OpenNebula developer community is driven by the company's internal business plan and hiring process. As a result, the proposed growth curves are not capable of describing the growth of the OpenNebula developer community.   

The \emph{OpenStack} project has a total number of 3800 developers, with 400 to 700 active contributors and 2000 to 6000 commit activities in each month. The community has a high diversity index of 5.60, and the Spearman's rank correlation coefficient between the number of active contributors and the number of commit activities is 0.98. The growth of the OpenStack developer community (Figure \ref{fig:subfig:growth_dev_openstack}) fits very well with the regression curves. The project spends significant amount of resources on persuading other organizations to join the ecosystem. As new organizations contribute developers to the project, they dilute the influence of the founding members. The number of active contributors does not depend on the business plan of any single organization, but on the collective resource invested on the project by many organizations. This is similar to the growth of bacteria in a nutrient-containing broth, with more nutrition being added to the broth continuously. When nutrition depletion is not an issue, the project remains in its exponential phase.

\section{Related Work}
\label{sec:related}

In recent years, open source software has received increasing level of attention from the academic community. Extensive literature exists on what open source software is \cite{weber2004success}, economical considerations such as intellectual property \cite{valimaki2005rise} and personal motivation \cite{hars2001working} \cite{hertel2003motivation}, how individual and organizations join and contribute to open source projects \cite{von2003community}, as well as the impact of community participation on the performance of commercial companies \cite{riehle2009commercial}. Most of these research papers have adopted a methodology that is qualitative rather than quantitative, and it is difficult to generalize the result. As a contrary, this paper focuses on quantitative analysis. The large amount of open source projects being included in this paper allows us to generalize the result.

Godfrey et al. \cite{godfrey2000evolution} investigated the growth and evolution of open source software, using the Linux kernel as a case study. They analyzed various releases of the Linux kernel source code between 1993 and 2001, using the gzip compressed size and lines of code to measure the growth of the Linux kernel project. They found that the Linux kernel project was growing in a linear fashion, which was surprising in that previously published research had suggested that large software systems tended to grow slower as they became larger. Robles et al. \cite{robles2005evolution} also studied the growth and evolution of both Linux kernel and *BSD kernel, as well as their sub-systems. They observed that most projects followed a linear but not smooth growth pattern, while super-linearity occurs only exceptionally. 

Mockus et al. \cite{mockus2002two} investigated the growth of the Apache web server project and the Mozilla web browser project, using email archives of source code change history and problem reports. They quantitatively studied aspects such as developer participation, core team size, code ownership, productivity, defect density, and bug resolution intervals for these two projects, and compared them with several commercial projects. The authors further proposed that a commercial/open source hybrid would yield in high-performance software development process (high productivity). Riehle et al. \cite{riehle2014paid} investigated the contribution from paid workers and volunteers in the Linux kernel and the Ohloh projects. They showed that about 50\% of all open source software development were paid work, and many small projects were fully paid for by companies. However, most important projects exhibited a balance in the amount of paid developer and volunteer work. The authors suggested that the ratio of volunteer to paid work could an indicator for the health / productivity of open source projects. In this paper, we reveal that the productivity of an open source project depends on the relationship between the contributing organizations, which is reflected in Spearman's rank correlation coefficient.

Oh et al. \cite{oh2007membership} investigated the interaction patterns among open source software community members, as well as the impact of size and connectivity on the stability of an open source software network. Their computer simulation results indicated that membership herding is negatively related to external influences and tends to occur in large networks with random connectivity. Huang et al. \cite{huang2011analysis} analyzed the structure and evolution of the Drupal content management system. The techniques they employed include social network analysis, degree distribution, hierarchical clustering, and scientific visualization. They reported that structure of the Drupal community displayed scale-free characteristics, which have been observed in a large number of other social, technical and biological networks. Colazo \cite{colazo2014structural} used Media Synchronicity Theory and Social Network Analysis to analyze the changes in the structure of collaboration networks for temporally dispersed collaborating teams. With archival data from 230 open source software projects, he concluded that the collaboration structure networks of more temporally dispersed teams are sparser and more centralized, and these associations are stronger in those teams exhibiting higher relative performance. These efforts provide a lot of insights into the structure of open source communities. However, they do not provide guidance on what kind of community structure is healthy, or how to achieve a healthy community structure. 

There exists a large volume of literature on the complex structure of community networks \cite{clauset2009power}\cite{palla2005uncovering}. Power law distribution is identified as a common characteristic of such communities. Mitzenmacher \cite{mitzenmacher2004brief} pointed out that communities in the area of computer science share this characteristic as well. Kumar et al. \cite{kumar2010structure}, Ahn et al. \cite{ahn2007analysis} reported results obeying the power law distribution from their studies on large scale online communities such as Flickr, Yahoo!'s 360, Cyworld, MySpace, and Orkut. The results presented in this paper conform very well with results in existing literature.

\section{Conclusion}
\label{sec:conclusion}

In this paper, we study the diversity, productivity, and growth of open source developer communities, using 20 open source projects in various disciplines as examples. We find out that (a) the Spearman's rank correlation coefficient between active contributors and commit activities reveals how changes in the size of the developer community impacts the productivity of the community; (b) the diversity index of an open source developer community reveals the structure of the community; and (c) the growth of open source developer communities can be described using growth curves such as the Gompertz curve and the logistic curve. In particular,

\begin{itemize}
	\item When the Spearman's rank correlation coefficient is high, the productivity of an open source developer community predictably increases when the size of the community grows. When the Spearman's rank correlation coefficient is low, the productivity of an open source developer community is not related to the size of the community.  
	\item For projects with low diversity, a dominating contributing organization controls the development plan. For projects with medium diversity, there exists a primary contributing organization and one or more secondary contributing organizations with significant contributions. For projects with high diversity, there no longer exists a primary contributing organization with a dominating contribution. Instead, there exists multiple major contributing organizations with approximately the same amount of code contributions. More importantly, the sum of the contributions from these major contributing organizations does not represent a dominating force among the developer community. 
    \item Projects with high diversity are more stable in that their Spearman's rank correlation coefficients tend to be high. The productivity of the community predictably increases when the size of the developer community grows. High diversity also contributes to the growth of the developer community in that changes in a number of the contributing organizations do not significantly impact the overall health of the community. As a result, communities with high diversity tend to stay in the exponential phase longer than other communities. To achieve long term growth of an open source developer community, it is undesirable to have a small number of contributing organizations dominating the community. Instead, it is desired to attract more contributing organizations to the community to achieve a greater diversity.
\end{itemize}

\bibliographystyle{IEEEtran}
\bibliography{open_source}

\end{document}